\documentclass{article}

\usepackage{PRIMEarxiv}

\usepackage[utf8]{inputenc} % allow utf-8 input
\usepackage[T1]{fontenc}    % use 8-bit T1 fonts
\usepackage{hyperref}       % hyperlinks
\usepackage{url}            % simple URL typesetting
\usepackage{booktabs}       % professional-quality tables
\usepackage{amsfonts}       % blackboard math symbols
\usepackage{nicefrac}       % compact symbols for 1/2, etc.
\usepackage{microtype}      % microtypography
\usepackage{lipsum}
\usepackage{fancyhdr}       % header
\usepackage{graphicx}       % graphics
\graphicspath{{media/}}     % organize your images and other figures under media/ folder

\title{Assessing the Performance of the DINOv2 Self-supervised Learning Vision Transformer Model for the Segmentation of the Left Atrium from MRI Images
}

\author{
  Bipasha Kundu, Bidur Khanal \\
  Center for Imaging Science Rochester \\
  Rochester Institute of Technology \\
  NY, USA\\
  \texttt{\{bk7944, bk9618\}rit@edu} \\
   \And
  Richard Simon, Cristian A. Linte \\
  Department of Biomedical Engineering \\
  Rochester Institute of Technology \\
  NY, USA\\
  \texttt{\{rasbme, calbme\}rit@edu} \\
}

\begin{document}
\maketitle

\begin{abstract}
% \lipsum[1]
Accurate segmentation of the left atrium from pre-operative scans is required for diagnosing atrial fibrillation, treatment planning, intraoperative guidance, and supporting computer-assisted surgical interventions. While deep learning models are pivotal in medical image segmentation, they often require extensive manually annotated datasets. However, the emergence of foundation models trained on larger datasets has helped to reduce this dependency, enhancing generalizability and robustness through transfer learning capabilities. In this work, we explore the out-of-the-box potential of DINOv2, a self-supervised learning vision transformer-based foundation model trained on natural images, by evaluating its performance in the left atrium (LA) segmentation task using MRI images. The challenges include the left atrium’s complex anatomical structures, thin myocardial walls, and limited annotated data, making it difficult to accurately segment the desired LA structures both prior to or during the image-guided intervention. We aim to demonstrate DINOv2’s ability to provide accurate and consistent segmentation in this specific context. We comprehensively evaluated the performance of DINOv2 in LA segmentation, utilizing end-to-end fine-tuning, and achieved a mean Dice score of 87.1\% and an Intersection over Union (IoU) of 79.2\%. Our study included data-level few-shot learning across different dataset sizes and patient counts, consistently finding that DINOv2 outperforms all baseline models. Furthermore, these comparisons suggest that DINOv2 can perform well out-of-the-box to match the above instances in the medical domain and effectively adapt and generalize to MRI data, even with minimal fine-tuning and limited data. These findings highlight DINOv2’s potential as a competitive tool for cardiac segmentation, providing accurate results essential for pre-procedural planning and pre-operative applications. Our study aims to inform medical researchers about DINOv2’s potential for broader implementation in other medical imaging modalities.
\end{abstract}

% keywords can be removed

\section{Introduction}
Atrial Fibrillation (AFib), a condition with irregular heart rhythm, is expected to affect 12 million people in the U.S. by 2030 \cite{into_1}. Accurate segmentation of the left atrium (LA) is important for diagnosing and treating AFib, as it helps identify the condition and guide interventional procedures like catheter ablation and the Maze procedure, both aiming to restore normal heart rhythm \cite{mcgann2014atrial}. Precise LA segmentation provides vital anatomical details, assisting surgeons in accurately targeting treatment areas, thereby enhancing the effectiveness of interventions and reducing complications. Additionally, it is essential for post-operative assessments, confirming the procedure's success, and monitoring for potential recurrences, ultimately ensuring better patient outcomes.
Recent advances in SSL have led to the development of powerful open-source AI models like DINOv2 \cite{oquab2023dinov2}, which have shown exceptional capabilities in zero-shot segmentation of natural images. Yet, significant differences exist between natural and medical image data, including variations in color, intensity, scaling, and anatomical structures \cite{into_3}. Medical domain data often present unique characteristics depending on the imaging
modality (CT, X-ray, or MRI). While experts can identify subtle changes and annotate these images accurately, deep learning models trained on natural images may perform less effectively in this domain. Given the challenges associated with collecting large annotated datasets comparable in size to those used in training DINOv2, it is worthwhile to investigate the potential of leveraging pre-trained DINOv2 models for medical image analysis, especially for segmentation tasks. This exploration could provide valuable insights into adapting such foundation models for specialized applications, bridging the gap between general-purpose AI and domain-specific demands.

The LA is adjacent to other anatomical structures with similar intensities to the blood pool, and its thin myocardial wall (2-3 mm) challenges imaging, even with high-resolution techniques. The scarcity of annotated data further complicates the segmentation process. \cite{tobon2015benchmark}. 
% The robustness of DINOv2, with its ability to provide accurate and generalizable segmentation, makes it an ideal candidate for tackling these challenges. 
We specifically chose the LA as a focus area in light of the segmentation challenges it poses caused by its variable behavior and the scarcity of extensive annotated datasets. 
Our study explores the potential of using DINOv2 to obtain a sufficiently accurate segmentation of the LA from MRI images, driven by the challenges posed by the complex and dynamic anatomical structure of the LA. By evaluating DINOv2 in this context, we aim to demonstrate its capability to deliver precise and consistent segmentation outcomes, even in complex and data-constrained scenarios.
We compared the performance of DINOv2 with state-of-the-art (SOTA) models such as Attention UNet (Att. UNet) \cite{oktay2018attention}, UNet \cite{ronneberger2015u}, and pre-trained ResNet50 backbone with UNet (Res50-UNet) \cite{sadad2021brain}. Additionally, we examine the performance of the model on data-level few-shot learning across various data percentages and patient counts. 
We chose to compare DINOv2 against Re50-UNet, UNet, and Attention UNet, despite their lower parameter counts and fully supervised training, to highlight how DINOv2, even with supervised fine-tuning on all data, excels in complex, data-constrained scenarios and successfully adapts from a natural domain to medical imaging.
By leveraging DINOv2's capability to focus on relevant features and adapt to new data, we aim to show its effectiveness in achieving accurate segmentation with minimal supervision.

\section{Methodology}
% \label{sec:headings}

% \lipsum[4] See Section \ref{sec:headings}.

\subsection{Segmentation with pre-trained DINOv2}
\begin{figure} [t]
   \begin{center}
   \begin{tabular}{c} 
   \includegraphics[height=5.7cm]{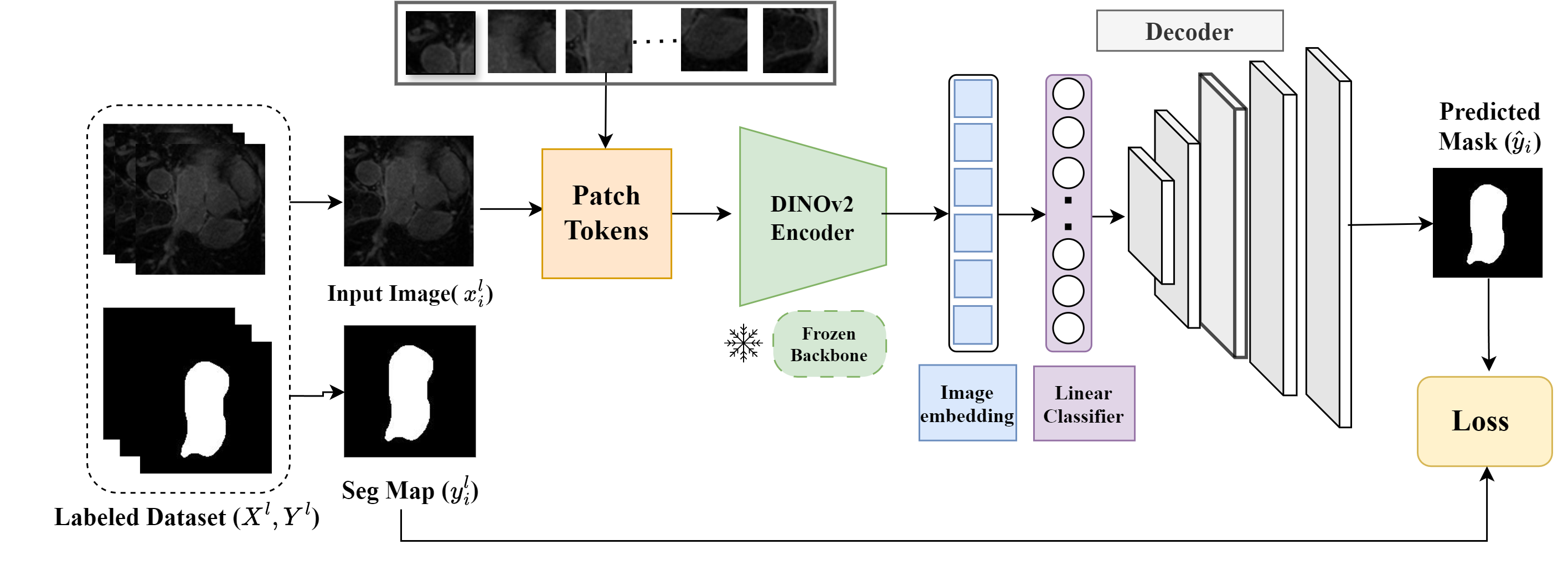}
	\end{tabular}
	\end{center}
   \caption[example] 
   { \label{fig_1}
   Detailed pipeline for fine-tuning DINOv2 on left atrium segmentation: utilizing labeled data to generate accurately predicted masks through transfer learning}
   \end{figure}

% \lipsum[5]
% \begin{equation}
% \xi _{ij}(t)=P(x_{t}=i,x_{t+1}=j|y,v,w;\theta)= {\frac {\alpha _{i}(t)a^{w_t}_{ij}\beta _{j}(t+1)b^{v_{t+1}}_{j}(y_{t+1})}{\sum _{i=1}^{N} \sum _{j=1}^{N} \alpha _{i}(t)a^{w_t}_{ij}\beta _{j}(t+1)b^{v_{t+1}}_{j}(y_{t+1})}}
% \end{equation}
DINOv2 \cite{oquab2023dinov2}, a state-of-the-art framework for SSL released by Meta in April 2023, is available on GitHub  \footnote {\url{https://github.com/facebookresearch/dinov2}}. It is trained on a diverse corpus of 142M curated natural images using various Vision Transformer (ViT) architectures. Our implementation utilizes the pre-trained DINOv2 model to extract robust features. It introduces a simple decoder designed to maintain resolution, in conjunction with the DINOv2's output processing layers, to segment the LA from cardiac MRI images. Monochromatic 2D slices of the cardiac LA are converted to 3-channel RGB images and resized to 448×448 pixels following the architecture requirements. We used the DINOv2 ViT-g/14 (giant) architecture, and its smaller versions, ViT-b/14 (base) and ViT-l/14 (large), with 1536, 768, and 1024 feature dimensions, respectively. We use the ViT-g/14 architecture to illustrate the segmentation workflow with DINOv2. Here, an input MRI slice (${x}_i$) is resized to 448×448 pixels and divided into (32$\times$32=)1024 non-overlapping patches of 14×14 pixels each, which are tokenized into 1536-dimensional tokens. A class token is concatenated with these patch tokens to encapsulate global semantic information. The DINOv2 model processes these tokens to extract corresponding image embeddings (feature representations), ${z}$, which are reshaped and permuted for convolutional processing. The LinearClassifierToken layer reshapes these features into a
format suitable for convolutional processing, consisting of a single convolutional layer with a kernel size of 1$\times$1, adjusting the number of channels and reshaping the feature map to 32×32. Then, the decoder, consisting of several convolutional and up-sampling layers, processes the reshaped feature maps to produce the final segmentation map ($\hat{y_i}$).
\begin{equation}    
  \centering  
 \hat{y_i} = Decoder(Reshape(DINOv2(x_{i})))
    % g_{seg}(x_i) = f_{UNet-Dec}{(f_{SSL}(x_i)})
\end{equation}
In our experiments, freezing the DINOv2 backbone leverages the robust features learned during pre-training, reducing computational load and focusing the learning process on the segmentation task. The detailed workflow of fine-tuning DINOv2 is illustrated in Fig. \ref{fig_1}

\subsection{Dataset Description}
% \lipsum[6]

% \paragraph{Paragraph}
% \lipsum[7]
The dataset was collected from the LAScarQs 2022 \cite{Lees-Miller-LaTeX-course-1} challenge hosted by MICCAI 2022. We evaluated our approaches to Task 2, focusing on the (semi)-automatic segmentation of the LA cavity from LGE MRI images. The dataset comprises 130 3D LGE MRI images with varying resolutions from 576×576 to 640×640 pixels and slice counts of either 88 or 44. All the data were acquired from AFib patients in a clinical setting, and gold standard labels for the LA cavity blood pool were also provided. For our study, we used 70\% of the 3D data for training, 10\% for validation, and 20\% for testing. We used 2D slices to train all our methods to leverage the detailed spatial information present in each slice for more effective segmentation.
\subsection{Pre- and Post-processing}

We ensured consistent pre- and post-processing for all methods (Attention UNet, UNet, Res50-UNet) except DINOv2. Pre-processing involved padding images to the highest resolution, resizing, cropping, and normalizing. Post-processing applied morphological operations \cite{post_processing}, specifically opening to remove artifacts and closing to smooth edges and fill holes, enhancing segmentation accuracy and quality. For fine-tuning DINOv2, we only performed normalization, with no additional pre- or post-processing.
\subsection{Implementation Details}
All techniques were implemented using the Pytorch framework, with experiments conducted on RIT's Research Computing Cluster equipped with NVIDIA A100 GPUs \cite{https://doi.org/10.34788/0s3g-qd15}. For the baseline methods chosen as the gold standard, input images were resized to 320$\times$320 and 448$\times$448 for all ViT architectures (ViT-b/14, ViT-l/14, and ViT-g/14). For the experiments, all models were trained using the Adam 
optimizer \cite{kingma2014adam} and BCEwithLogits loss \cite{loss_fn} with a learning rate of 0.001 for DINOv2 and 0.0001 for all other methods. The training setup included 75 epochs and a batch size of 24 for all baseline methods except DINOv2, which was trained for 35 epochs with a batch size of 32. The choice of larger batch size for DINOv2 is primarily motivated by the distinct characteristics and computational demands of ViT architectures compared to traditional CNN-based baseline models. We used early stopping to avoid overfitting, and the best validation checkpoints were selected for testing.

\section{Preliminary Results}
% \label{sec:others}
% \lipsum[8] \cite{kour2014real,kour2014fast} and see \cite{hadash2018estimate}.

% The documentation for \verb+natbib+ may be found at

We conducted experiments using fully supervised fine-tuning and data-level few-shot learning approaches to assess DINOv2's adaptability to the LA data. We evaluated our experiments using the dice score and Jaccard index (IoU), including standard deviations (SD). The methods assessed include Att. UNet, UNet, Res50-UNet, and different versions of DINOv2. Table \ref{tab_dice} shows the quantitative comparison of evaluation metrics after fine-tuning DINOv2 for segmentation alongside SOTA models Att. UNet, UNet, and Res50-UNet. Our results indicate that higher scores indicate better performance, which is noted with ↑. 
%#############################################################
 \begin{table}[!ht]
   \caption{Quantitative Comparison of Dice Score (\%) and IoU (\%) with standard deviation for left atrium segmentation} 
   \vspace{.25cm}
   \label{tab_dice}
   \small
   \centering
   \setlength{\tabcolsep}{12pt} % Adjust column separation here
   \renewcommand{\arraystretch}{1.25} % Adjust row separation here
   % \resizebox{\textwidth}{13}{ % Adjust the table to the full width of the text
   \begin{tabular}{lccccrr}
   \toprule
   \textbf{Methods} & \textbf{Dice  $\uparrow$} & \textbf{IoU $\uparrow$} \\
   \midrule
   Attention UNet& 79.2 ± 12.3 & 72.1 ± 13.8 \\
   UNet & 84.1 ± 8.3   &   76.4 ± 12.8   \\
  Pre-trained ResNet50-UNet& 84.6 ± 13.1 & 76.2 ± 12.1 \\
  
  DINOv2 ViT-base & 84.9 ± 7.2 & 75.5 ± 6.3  \\        
   DINOv2 ViT-large  & 85.6 ± 5.1 & 77.0 ± 6.0    \\
    \textbf{DINOv2 ViT-giant}  & \textbf{87.1 ± 4.8} & \textbf{79.2 ± 5.2} \\
   %  V2S-Net &  2.49±1.18 & 4.17±1.38   \\
   % *GMM-FEM  & \textbf{ 1.32±1.06} & \textbf{ 2.18±1.03}  \\
   \bottomrule
   \end{tabular}
\end{table}
% #########################################################
% \begin{center}
%   \url{http://mirrors.ctan.org/macros/latex/contrib/natbib/natnotes.pdf}
% \end{center}
% Of note is the command \verb+\citet+, which produces citations
% appropriate for use in inline text.  For example,
% \begin{verbatim}
%    \citet{hasselmo} investigated\dots
% \end{verbatim}
% produces
% \begin{quote}
%   Hasselmo, et al.\ (1995) investigated\dots
% \end{quote}

\begin{figure}[ht]
   \begin{center}
   \begin{tabular}{c} %% tabular useful for creating an array of images 
   \includegraphics[height=5.8cm]{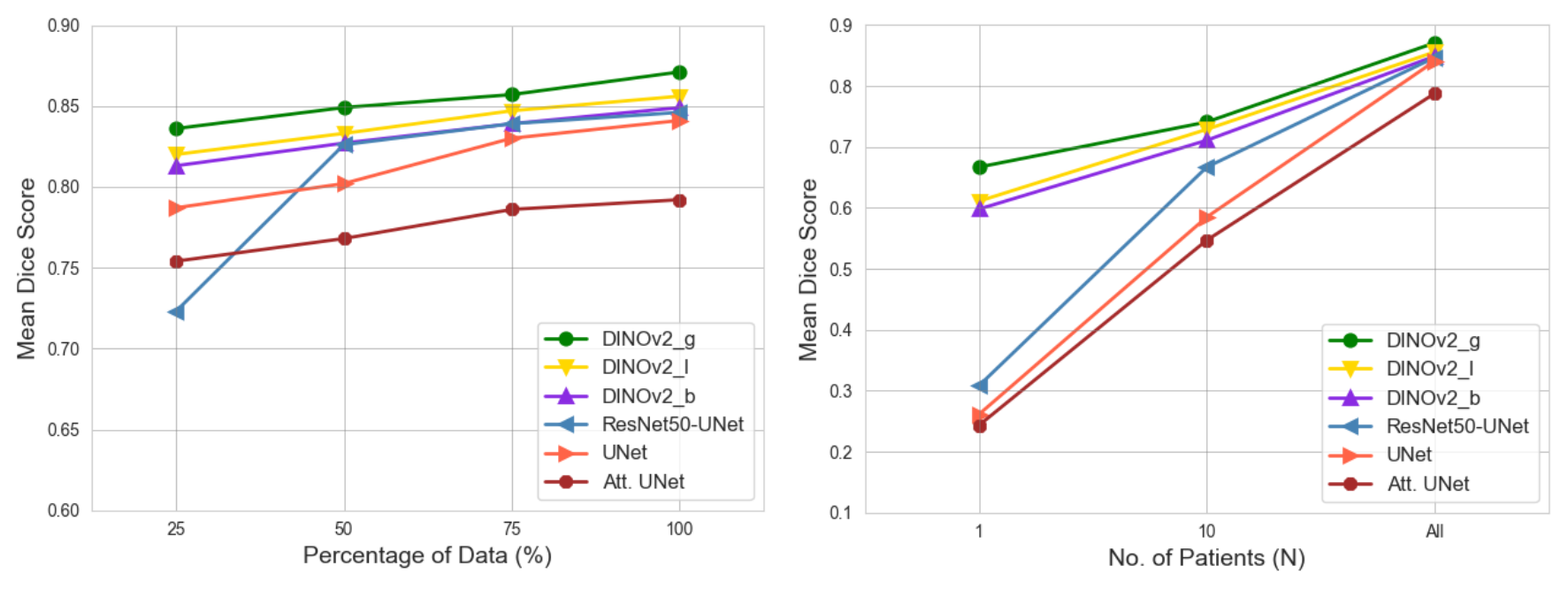}
   \end{tabular}
   \end{center}
   \caption[example] 
%>>>> use \label inside caption to get Fig. number with \ref{}
   { \label{fig_2} 
Comparative analysis of data level few-shot learning performance across all methods: Left - evaluation of performance metrics with varying dataset sizes; Right -  evaluation of performance metric with different patient counts}
   \end{figure}

% \begin{center}
%   \url{https://www.ctan.org/pkg/booktabs}
% \end{center}
The results show that UNet outperformed Att. UNet, achieving a Dice Score of 84.1\% compared to 79.2\%.  The Res50-UNet trained on ImageNet data further enhanced performance close to DINOv2-base architecture, leveraging transfer learning from its pre-trained encoder, and achieved a Dice Score of 84.6\%. Among the DINOv2 models, the ViT-base achieved a Dice Score of 84.9\%, while ViT-large showed an improvement with a Dice Score of 85.6\%. The ViT-giant version of DINOv2 demonstrated the best performance, with a Dice Score of 87.1\%.
We also performed data-level few-shot learning with different patient counts and data sizes, with minimal fine-tuning \& testing. As shown in Fig. \ref{fig_2}, the left subplot demonstrates that all DINOv2 architectures consistently outperform other methods across different data sizes, particularly when the data sizes are reduced. The right subplot highlights DINOv2's superior performance when trained on data from one, ten, and all patients. It clearly demonstrates its robustness and efficiency, particularly in few-shot and limited-data scenarios.

\begin{figure} [!ht]
   \begin{center}
   \begin{tabular}{c} %% tabular useful for creating an array of images 
   \includegraphics[height=9cm]{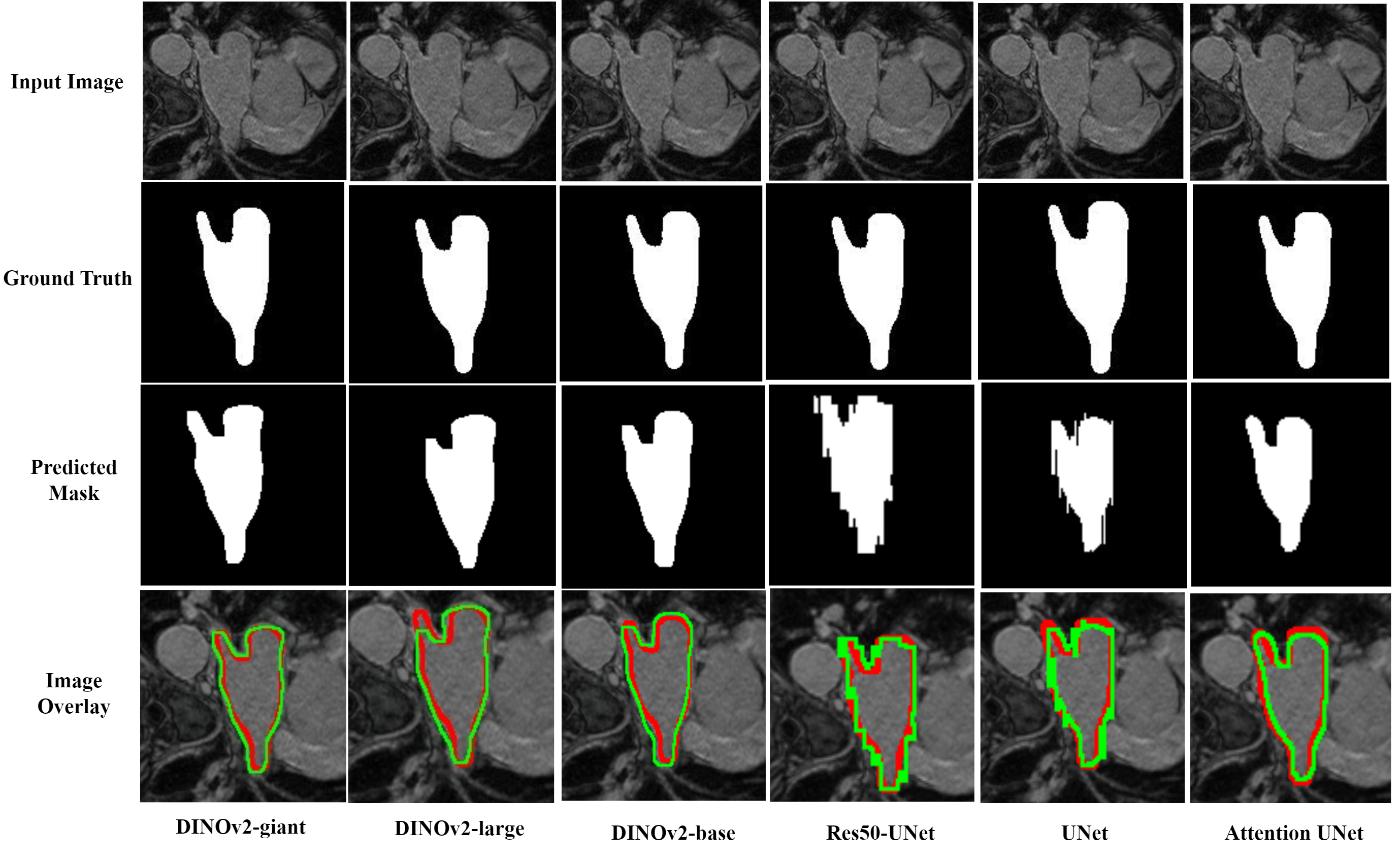}
   \end{tabular}
   \end{center}
   \caption[example] 
%>>>> use \label inside caption to get Fig. number with \ref{}
   { \label{fig_3} 
Qualitative comparison of binary segmentation results: Overlaying predictions (green) and ground truth (red) on input images for left atrium segmentation}
   \end{figure}

We also provide a qualitative comparison among different DINOv2 architectures and the SOTA Attention UNet, UNet, and Res50-UNet in Fig. \ref{fig_3}. Notably, DINOv2 performs better in delineating the complex structures of the LA with clearer boundaries, fine anatomical details, and less noise than other traditional CNN-based methods. The qualitative results show that baseline methods struggle with some edge cases, leading to less precise segmentation than the foundation model. This visual assessment underscores the advantage of using a foundation model like DINOv2 for medical image segmentation tasks, providing both better accuracy and consistency, especially in cases with complex anatomical variations.

\section{New Or Breakthrough Work To Be Presented}
% NEW OR BREAKTHROUGH WORK TO BE PRESENTED 
Considering the challenges (complex anatomical structures, thin wall boundaries, and limited annotated data) associated with the left atrium segmentation, our study leverages the pre-trained capabilities of DINOv2 to provide a more accurate analysis of LA segmentation from MRI images. Our work underscores the out-of-the-box potential of the large foundation model, DINOv2, in efficiently handling complex medical data, even when they are initially trained on natural domain images. Our preliminary results highlight the robustness, scalability, effectiveness, and generalizability of these DINOv2 models (base, large, giant) in the medical domain, setting a new benchmark for segmentation tasks and encouraging further exploration of these open-source foundation models in other medical imaging fields.

% \subsection{Figures}
% \lipsum[10] 
% See Figure \ref{fig:fig1}. Here is how you add footnotes. \footnote{Sample of the first footnote.}
% \lipsum[11] 

% \begin{figure}
%   \centering
%   \fbox{\rule[-.5cm]{4cm}{4cm} \rule[-.5cm]{4cm}{0cm}}
%   \caption{Sample figure caption.}
%   \label{fig:fig1}
% \end{figure}

% \subsection{Tables}
% \lipsum[12]
% See awesome Table~\ref{tab:table}.

% \begin{table}
%  \caption{Sample table title}
%   \centering
%   \begin{tabular}{lll}
%     \toprule
%     \multicolumn{2}{c}{Part}                   \\
%     \cmidrule(r){1-2}
%     Name     & Description     & Size ($\mu$m) \\
%     \midrule
%     Dendrite & Input terminal  & $\sim$100     \\
%     Axon     & Output terminal & $\sim$10      \\
%     Soma     & Cell body       & up to $10^6$  \\
%     \bottomrule
%   \end{tabular}
%   \label{tab:table}
% \end{table}

% \subsection{Lists}
% \begin{itemize}
% \item Lorem ipsum dolor sit amet
% \item consectetur adipiscing elit. 
% \item Aliquam dignissim blandit est, in dictum tortor gravida eget. In ac rutrum magna.
% \end{itemize}

\section{Conclusion}
In this paper, we explored the out-of-the-box potential of DINOv2 as a foundation model for LA segmentation. We evaluated the robustness and performance of DINOv2 using both end-to-end fine-tuning and few-shot learning approaches for varying data sizes and patient counts. Our preliminary results indicate that all versions of DINOv2 outperform with a higher dice score, especially excelling with less data. This highlights a trade-off: DINOv2 utilizes advanced features to perform well with less data, while the baseline models may require more data to achieve similar results. The reasonably low standard deviation across the DINOv2 models shows a consistent performance, whereas we found some traditional models to be subpar, mainly when dealing with limited data. We will include three additional baseline methods, nnUNet, pre-trained Res152-UNet, and SegNet, and we will conduct a comprehensive statistical \& time complexity analysis of all methods in the upcoming conference proceedings manuscript. We also intend to include more patient counts for the few-shot learning approach.  This study underscores the value of leveraging open-source foundation models like DINOv2, pre-trained on large, diverse natural image datasets that can learn rich and transferable features for specific applications such as segmentation of left atrium. The use of these models can lead to enhanced accuracy and robustness in segmentation tasks, making them valuable choices for broader medical image analysis applications.

% \section*{Acknowledgments}
% This was was supported in part by......

%Bibliography
\bibliographystyle{unsrt}  
\bibliography{references}

\end{document}